\documentclass[onecolumn]{elsarticle}

\usepackage{url}
\usepackage[a4paper,margin=1in]{geometry}
\usepackage{graphicx}
\usepackage{float}
\usepackage{enumitem}
\usepackage{comment}
\usepackage{amssymb}
\usepackage[caption = false]{subfig}
\usepackage[table,xcdraw]{xcolor}
\usepackage{float}
\usepackage{mathptmx}
\newcommand{\sars}{SARS-CoV-2~}
\newcommand{\covid}{COVID-19~}
\usepackage{lineno}
\usepackage{caption}
\usepackage[caption = false]{subfig}

\journal{Journal Name}

\usepackage{lipsum}
\makeatletter
\def\ps@pprintTitle{%
 \let\@oddhead\@empty
 \let\@evenhead\@empty
 \def\@oddfoot{}%
 \let\@evenfoot\@oddfoot}
\makeatother

\begin{document}

\begin{frontmatter}

\title{Analyzing Host-Viral Interactome of SARS-CoV-2 for Identifying Vulnerable Host Proteins during COVID-19 Pathogenesis}



\author{Jayanta Kumar Das\fnref{label1}}
\ead{jdas4@jhmi.edu}
\author{Swarup Roy \corref{cor1}\fnref{label2}}
\ead{sroy01@cus.ac.in}
\author{Pietro Hiram Guzzi \corref{cor1}\fnref{label3}}
\ead{sroy01@cus.ac.in}
\cortext[cor1]{Corresponding Author}
\address[label1]{Department of Pediatrics, Johns Hopkins University School of Medicine, Maryland, USA}
\address[label2]{Network Reconstruction \& Analysis (NetRA) Lab, Department of Computer Applications, Sikkim University, Gangtok, India}
\address[label3]{Department of Medical and Surgical Sciences, Magna Graecia University, Catanzaro, Italy}
\author{}

\address{}

\begin{abstract}

The development of therapeutic targets for \covid treatment is based on the understanding of the molecular mechanism of pathogenesis. The identification of genes and proteins involved in the infection mechanism is the key to shed out light into the complex molecular mechanisms. The combined effort of many laboratories distributed throughout the world has produced the accumulation of both protein and genetic interactions. In this work we integrate these available results and we obtain an host protein-protein interaction network composed by 1432 human  proteins. We calculate network centrality measures to identify key proteins. Then we perform functional enrichment of central proteins. We observed that the identified proteins are mostly associated with several crucial pathways, including cellular process, signalling transduction, neurodegenerative disease. Finally, we focused on proteins involved in causing disease in the human respiratory tract. We conclude that \covid is a complex disease, and we highlighted many potential therapeutic targets including RBX1, HSPA5, ITCH, RAB7A, RAB5A, RAB8A, PSMC5, CAPZB, CANX, IGF2R, HSPA1A, which are central and also associated with multiple diseases.

\end{abstract}

\end{frontmatter}

\section{Introduction}
The world is experiencing an unprecedented pandemic due to a massive outbreak of Severe Acute Respiratory Syndrome Corona Virus 2 (\sars) infected viral disease, COVID-19. \sars, is a large enveloped coronavirus (family-\emph{Coronaviridae},  subfamily-\emph{Coronavirinae}) with non-segmented, single-stranded, and positive-sense RNA genomes~\cite{wrapp2020cryo}, transmits rapidly through human to human contacts. Although \sars is similar to other known coronaviruses, i.e. SARS-CoV and MERS-CoV~\cite{perlman2009coronaviruses,de2013commentary}, it has demonstrated high rates of infection \cite{liu2020reproductive,surveillances2020epidemiological}. Therefore there is the need to understand the disease pathogenesis of SAR-CoV-2 to develop effective therapies and vaccines.

The \sars virus is responsible \covid disease that causes damages in multiple organs as the disease progresses from an asymptomatic phase to a life-threatening disease~\cite{servick2020survivors}. Therefore, accurate molecular diagnosis of COVID-19 disease is essential by collecting the proper respiratory tract specimen~\cite{whetton2020proteomics}. In this context, the integrated analysis \cite{antonelli2019integrating} of various data-sets, including clinical and imaging data, may explain, and hopefully predict,  the longitudinal effects of SARS-CoV-2 infection~\cite{tang2020laboratory,dasbib}. In particular, many independent projects throughout the world have focused on genomics and proteomics level~\cite{dasbib}, and then they integrated these data with clinical ones. 
These works have produced data about the infection's effect at a molecular scale, evidencing genes and proteins' role, such as the interactions among viral and human proteins. 
Interactions between a host and its pathogen, are primarily driven by interactions among the host proteins and pathogen proteins; also referred to as host-pathogen protein-protein interaction (PPI) network. The \sars virus-host interactome have been studied focusing various virulence factors influencing SARS-CoV-2 pathogenesis and interacting mechanism~\cite{guzzi2020master,hoffmann2020functional,messina2020covid,li2020virus,cannataro2010impreco,agapito2013cloud4snp}. Further, many recent works also used host-viral protein-protein interaction network as an input to elucidate potential drug targets or repurposed drug molecules~\cite{gordon2020sars,zhou2020network,beck2020predicting}. Host-pathogen protein interactions provide important insights into the molecular mechanisms of pathogenecity~\cite{memivsevic2015mining} and for understanding virulence factors influencing \sars pathogenesis~\cite{liang2020virus,thiel2003mechanisms}. \sars is a newly found virus whose interacting human host proteins play a major disease progression role that needs to be investigated.

\begin{figure*}[ht!]
\centering
\includegraphics[width=\textwidth]{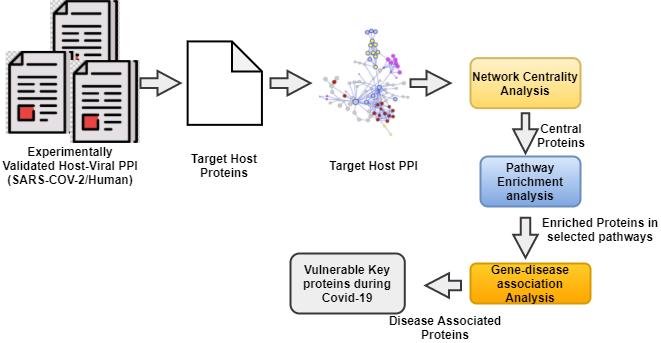}
\caption{The complete work-flow of the current study.}
\label{fig:work-flow}
\end{figure*}

Protein-Protein Interactions (PPI) are usually modelled and analysed with graph theory \cite{guzzi2020biological}. In this formalism, the interactions are modelled as a graph whose nodes are proteins (or genes), and the edges are the interaction among them. Several studies have found that specific candidate proteins might play a crucial role ~\cite{li2013prioritizing,ferrari2018stratification,galicia2020predicting,lim2011identification}. Protein-protein interaction networks are an essential ingredient for any systems-level understanding of cellular processes and modelling, and even drug discovery~\cite{tucker2001towards,thakur2015review,athanasios2017protein,chautard2009interaction,nietzsche2016protein}. The key genes/proteins involved in the different biological pathways can give valuable insight for in-depth characterisation of disease progression~\cite{lan2015computational,safari2014protein,wang2018integrated}. It is well accepted that all the viruses have evolved to target proteins that are central and have strong control over the human interactome~\cite{jeong2001lethality,bosl2019common,albert2000error,navratil2011human,halehalli2015molecular}. Exploring the predicted interaction networks can suggest new directions for future experimental research and provide cross-species predictions for efficient interaction mapping~\cite{xu2006discovering,safari2014protein}. The complete workflow of the current study can be seen from Figure~\ref{fig:work-flow}.

This study aims to identify essential human host proteins based on topology analysis of the protein-protein interaction network of \sars interacting human host proteins. We performed functional enrichment of the identified proteins to shed out light on cellular, signalling, and disease pathways. 

\section{Materials and Method} \label{method}

\subsection{Dataset: Curated \sars interacting human host proteins}

\begin{figure*}[ht!]
\centering
\subfloat[] {\includegraphics[width=7cm,height=6cm]{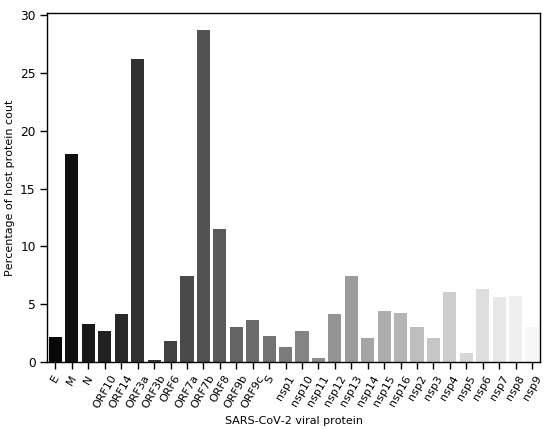} } \quad
\subfloat[] {\includegraphics[width=7cm,height=6cm]{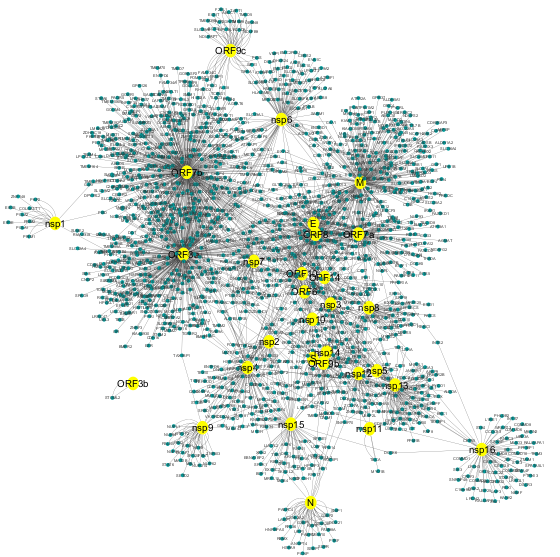}}
\caption{The abundance (percentage) of collected interacting human host protein for different \sars  viral proteins. A host-viral interaction network pattern is also shown.}
\label{fig:data}
\end{figure*}

We use recently reported host proteins that are physically verified using Affinity purification mass spectrometry for their interactions with \sars~\cite{gordon2020sars,liang2020virus,stukalov2020multi}. The used host-viral protein interactions are also available in  BioGRID ~\cite{stark2006biogrid}. A total of 2489 host-viral interactions (consisting of 1432 unique host proteins interacting with 37 \sars viral proteins) are obtained. In Figure~\ref{fig:data}, we provided the number of interacting host protein count. It is noted that the majority of the host proteins are targeted to the specific viral protein. 

\subsection{Construction of Host PPI network}


Starting with the human proteins that are interacting with the virus, we build a host PPI by querying the Search Tool for the Retrieval of Interacting Genes/Proteins (STRING, Version 10.0;  \url{http://string-db.org/})~\cite{szklarczyk2010string}.


The topology analysis of the PPI network is performed by using Cytoscape  (\url{http://apps.cytoscape.org}), a general platform for complex network analysis and visualization~\cite{shannon2003cytoscape}.

\subsection{Centrality analysis of host PPI network}
\label{sec:centraity}
In network analysis, indicators of centrality identify the most critical nodes in the network~\cite{bonacich1987power}. The centrality measure uses to characterise each node and edge in the PPI network. The degree measure is the most intuitive for topology analysis of the PPI network. Several other crucial factors that can influence network links are betweenness centrality, closeness centrality, clustering coefficient, topological coefficient, and neighbourhood connectivity.

\begin{enumerate}[label=(\roman*)]
\item \textbf{Degree centrality:} The degree centrality (simply degree) of a node $n$ in a network is defined  as ($D_n$), which indicates number of directly connected nodes to $n$. The densely connected nodes in PPI network is considered hub nodes~\cite{han2004evidence}.

\item \textbf{Betweenness centrality:} Betweenness centrality quantifies the number of times a node acts as a bridge along the shortest path between two other nodes~\cite{yoon2006algorithm}. The betweenness centrality of a node $n$ is represented as:
\begin{equation}
  C_b(n)=\sum_{s\neq n \neq t}(\sigma_{st(n)}/\sigma_{st})
\end{equation}
where $\sigma_{st}$ is the total number of shortest paths from node $s$ to node $t$ and $\sigma_{st}(n)$ is the number of those paths that pass through $n$.

\item \textbf{Closeness centrality:} Closeness centrality is a way of detecting nodes that are able to spread information very efficiently through the network~\cite{newman2005measure}. It can be calculated as : 
\begin{equation}
C_c(n)=1/avg(L(m,n))
\end{equation}
where $L(m,n)$ is the length of shortest path between node $n$ and $m$, and $m$ denotes any other nodes that are reachable to node $n$.

\item \textbf{Average shortest-path length:} Shortest-path length between two nodes (say $n$ and $m$) in network topology is defined as the number of minimum steps that required to traverse between node $n$ and $m$!\cite{mao2013analysis}. The average shortest path length ($S_p$) of node $n$ is the average value of all pair of nodes shortest path from the node $n$. 

\item \textbf{Clustering coefficient}: Clustering coefficient is a measure of the degree to which nodes in a graph tend to cluster together~\cite{barabasi2004network}. In undirected networks, the clustering coefficient $C_n$ of a node $n$ is defined as: 
\begin{equation}
    C_n = 2e_n/(k_n(k_n-1))
\end{equation}
where $k_n$ is the number of neighbors of $n$ and $e_n$ is the number of connected pairs between all neighbors of $n$.

\item \textbf{Topological coefficient:} Topological coefficient is a relative measure for the extent to which a node shares neighbors with other nodes~\cite{goldberg2003assessing}. The topological coefficient $T_n$ of a node $n$ with $k_n$ neighbors is computed as follows:
\begin{equation}
T_n = avg ( J(n,m) ) / k_n
\end{equation}

Where $J(n,m)$ is defined for all nodes $m$ that share at least one neighbour with $n$, and the value $J(n,m)$ is the number of neighbours shared between the nodes $n$ and $m$, plus one if there is a direct link between $n$ and $m$.

\item \textbf{Neighborhood connectivity:} Neighborhood connectivity ($N_c$) of a node $n$ is defined as the average connectivity of all neighbors of $n$~\cite{maslov2002specificity}. The neighborhood connectivity distribution gives the average of the neighborhood connectivities of all nodes $n$ with $k$ neighbors for $k = 0,1,\cdots$.
\end{enumerate}

We used NetworkAnalyzer~\cite{shannon2003cytoscape} to calculate above centrality score. In NetworkAnalyzer, $C_c$ (Closeness centrality) is calculated as the reciprocal of the average shortest path length. So, high $C_c$ means highly central, and thus low $S_p$.

\subsection{Gene ontology and pathway enrichment analysis}
We performed enrichment analysis to find out set of significantly enriched genes/proteins in different functional and biological pathways. We used KEGG (Kyoto Encyclopedia of Genes and Genomes) ~\cite{kanehisa2000kegg} for elucidating pathway enrichment of a host protein and Gene Ontology (GO) for the assessment of protein functions~\cite{ashburner2000gene}. KEGG is a database resource for understanding high-level functions and utilities of the biological system~\cite{kuleshov2016enrichr}. 

\subsection{Gene-disease association network}

Complex diseases are caused by a group of genes known as disease genes. More often, a gene can participate in various disease conditions~\cite{goh2007human,wellcome2007genome}. It helps unravel the disease pathogenesis, which in turn help disease diagnosis, treatment, and disease prevention. We obtained gene-disease association network from DisGeNET (v7.0) database (\url{https://www.disgenet.org/}), which contains 1,134,942 gene-disease associations (GDAs), between 21,671 genes and 30,170 disease ~\cite{pinero2020disgenet}. From this database, we considered curated gene-disease associations only.

\section{Results and Discussion}
Here, we report the outcomes of intermediate steps to reach to our objective of isolating key host proteins followed by their significance analysis. 
\subsection{Deriving PPI network for candidate host proteins}
\begin{figure*}[h!]
\centering
\includegraphics[scale=.95]{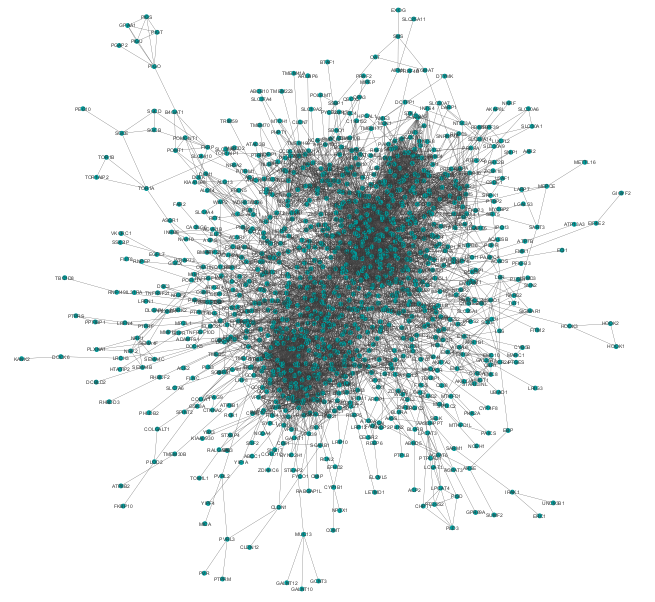}
\caption{The gain component of PPI network consisting of 1111 nodes and 7043 edges obtained from whole PPI network.}
\label{fig:ppi}
\end{figure*}

Our candidate host proteins list, collected from the reported host-viral networks (Section~\ref{method}), consists of total of 1432 distinct proteins that are targeted by \sars during \covid. We rebuilt the PPI network centered around our candidate proteins using STRING DB. There are 7076 edges in the derived PPI network. We curated derived PPI by keeping only the interactions whose confidence scores are at least $0.7$ (high confidence). The derived PPI network is then analysed using Cytoscape. We identified the big connected component (also called gain/main component) of the PPI network. After discarding all disconnected components in the PPI network, we considered gain component of PPI network with 1111 nodes (Approx. $78\%$) and 7043 edges (Figure~\ref{fig:ppi}).

\subsection{Network topology analysis of gain component}

\begin{figure}[!ht]
\centering
\includegraphics[width=.6\textwidth]{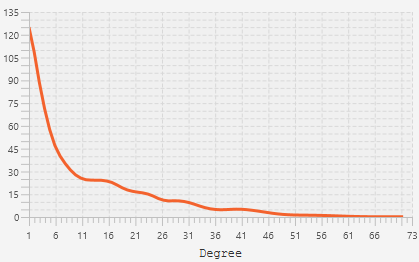}
\caption{The degree distribution of all 1111 nodes (proteins) in gain component of PPI network. The x-axis indicates degree distribution, whereas y-axis shows relative frequency distributions.}
\label{fig:degree} 
\end{figure}

 We performed topological analysis of the gain component using NetworkAnalyzer~\cite{shannon2003cytoscape}. The degree distribution of all the candidate proteins in the gain component showed that the majority of the proteins in the gain component exhibit a higher degree of connectivity (Figure~\ref{fig:degree}). Few proteins with degree (shown within parentheses) more than 50 are \emph{CDK1(73), PPP2R1A(65), NOP56(60), POLR2B(60), RAB1A(59), RBX1(58), SKIV2L2(57), NAPA(57), RPS14(56), STX5(54), TGOLN2(54), TCEB1(53), DCTN2(53), TCEB2(52), HSPA9(51), GNB2L1(50)}.

The histogram analysis of all the centrality measures (discussed in section~\ref{sec:centraity}) showing non-random distribution (Figure S1). We performed correlation (Pearson) analysis among all centrality scores (Table~\ref{tab:corr}). The correlation score between degree centrality ($D_c$) scores and closeness centrality ($C_c$) scores observed to be the highest ($r=0.759$) in comparison to other measures. Although, we observed correlation between $D_C$ and neighbourhood centrality ($N_{c}$) is the third-highest ($r=0.557$), but $N_c$ and $B_c$ showed less correlative ($r=0.1101$). Overall, we observed correlation score of three centrality measures ($D_C, B_c,C_c$) are quite closer. Therefore, we selected them in subsequent analysis. We identified 373 proteins in these criteria, which are considered highly central proteins (above the median score for all three selected parameters). When we considered all measures, we find only six common proteins (\emph{GEMIN4, DDX20, GOLGA3, FKBP15, PMPCA, AK4}) above the median score in each category of centrality measurement, and that is the reason why we selected three centrality measures for our downstream analysis.

\begin{table}[h!]
\centering
\caption{Correlation analysis among all centrality parameters computed for 1111 proteins (Figure~\ref{fig:ppi}).} \label{tab:corr}
\begin{tabular}{l|lllll}  
   & $D_c$     & $B_c$      & $C_{coef}$     & $T_c$      & $N_c$     \\ \hline 
$B_c$  & \textbf{0.603} & 1      &       &        &       \\
$C_{coef}$  & 0.209 & -0.168 & 1     &        &       \\
$T_c$  & -0.32 & -0.283 & 0.451 & 1      &       \\
$N_c$  & 0.557 & 0.1101 & 0.346 & -0.139 & 1     \\
$C_c$ & \textbf{0.759} & \textbf{0.5146} & 0.213 & -0.329 & 0.684 \\
\hline 
\end{tabular}
\end{table}

\begin{figure*}[ht!]
\centering
\includegraphics[width=\textwidth]{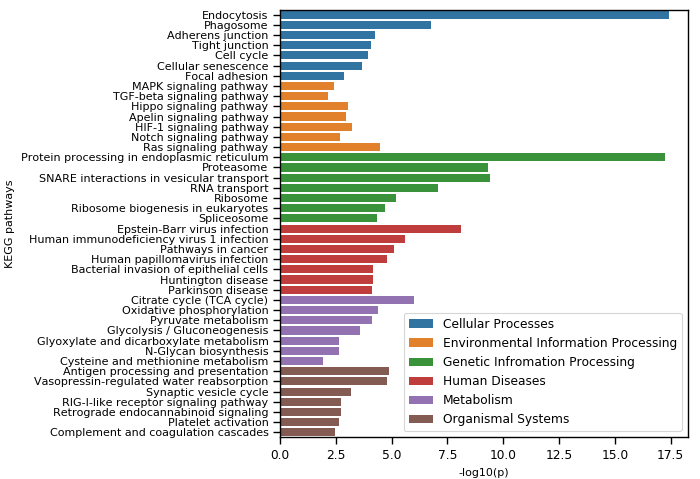}
\caption{The top 7 enriched pathways in each category of KEGG pathways. In each category, pathways are shown ordered by $-log10(p)$ value..}
\label{fig:pathway7}
\end{figure*}

\begin{figure*}[ht!]
\centering
\includegraphics[scale=.85]{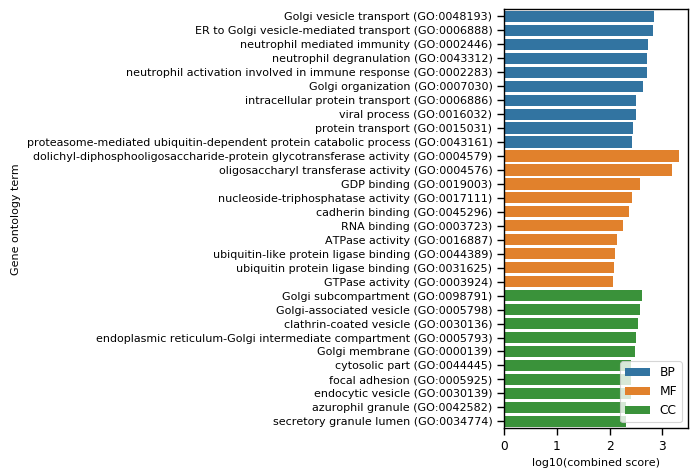}
\caption{The top 10 enriched terms in each category of gene ontology (BP-Biological process, MF-Molecular function, CC-Cellular component). In each category, terms are shown ordered by $log10(combined score)$ value..}
\label{fig:go10}
\end{figure*}

\subsection{Pathway enrichment analysis of highly central proteins}
We performed KEGG pathway analysis of selected 373 highly central proteins. We obtained a total of 84 enriched KEGG pathways within the significant level ($adj-p<.05$). The enriched pathways were involved in Cellular Processes (9), Environmental Information Processing (9), Genetic Information Processing (13), Human Disease (31), Organismal Systems (15), Metabolism (7). The top seven pathways in each category are shown in Figure~\ref{fig:pathway7}.

Our current study mainly focused on the proteins that are involved in cellular process, signalling transduction, and human disease (viral and neurodegenerative) pathways, the most affected pathways in the context of \covid disease~\cite{seif2020jak,ganesan2019mtor,luo2020targeting,grimes2020p38}. There are  nine enriched pathways in \textbf{cellular process} (Endocytosis, Phagosome, Adherens junction, Tight junction, Cell cycle, Cellular senescence, Focal adhesion, Regulation of actin cytoskeleton, Lysosome), nine pathways in \textbf{Environmental Information Processing}-signalling transduction (Ras signalling pathway, HIF-1 signalling pathway, Hippo signalling pathway, Apelin signalling pathway, MAPK signalling pathway, TGF-beta signalling pathway, AMPK signalling pathway, NF-kappa B signalling pathway), nine pathways from \textbf{human disease viral} sub-category (Human immunodeficiency virus 1 infection, Human papillomavirus infection, Human cytomegalovirus infection, Hepatitis B, Human T-cell leukaemia virus 1 infection, Influenza A, Hepatitis C, Measles) and four pathways from \textbf{ neurodegenerative disease } with sub-category (Huntington disease, Parkinson disease, Alzheimer disease, Prion diseases). A total of 141 distinct proteins (out of 373) were obtained from these pathways, which are then ranked based on presence in selected enriched pathways, and we found that 79 proteins are associated in our candidate pathways. All these proteins were then further studied for disease-gene association in the next.

We also performed gene set enrichment analysis (Gene ontology).  It is observed that out of selected genes mostly involved in Biological process (\emph{Supplementary-B}). The top ten terms in each category of gene ontology (BP, MF, CC) are shown in Figure~\ref{fig:go10} that includes 
neutrophil mediated immunity (GO:0002446),
neutrophil activation involved in immune response (GO:0002283) and viral process (GO:0016032) from BP category; dolichyl-diphosphooligosaccharide-protein glycotransferase activity (GO:0004579), GDP binding (GO:0019003), 
cadherin binding (GO:0045296) and 
ATPase activity (GO:0016887) from MF category; and focal adhesion (GO:0005925) from CC category.

\begin{figure*}[ht!]
\centering
\subfloat[Gene comparison] {\includegraphics[width=7cm,height=6cm]{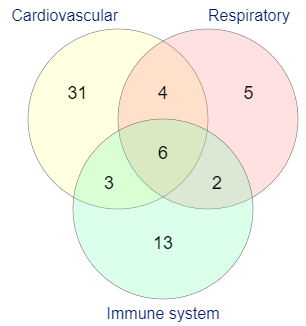} } \quad
\subfloat[Disease comparison] {\includegraphics[width=7cm,height=6cm]{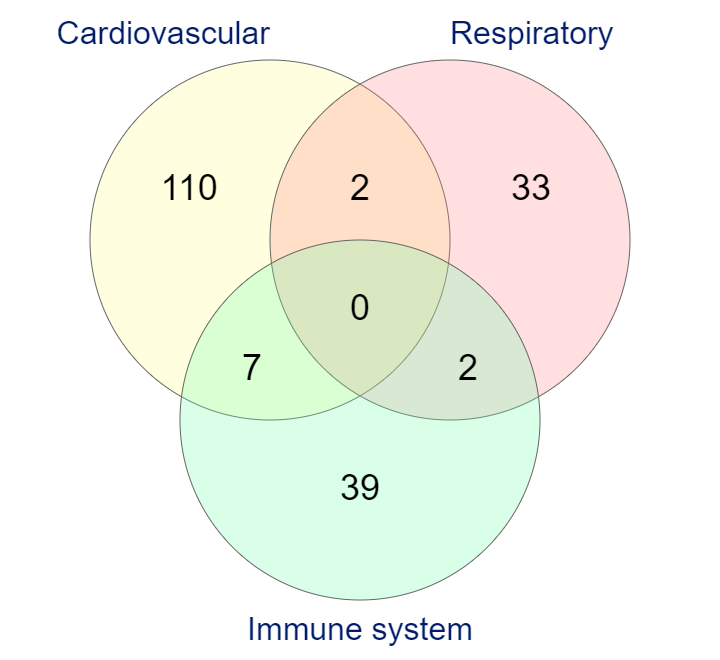}}
\caption{Comparison of three groups of disease categories (Cardiovascular, Respiratory, Immune system) using venn-diagram. (a) based on number protein count in each category; (b) based on number of disease associated (curated from database) among the observed protein in each category.}
\label{fig:ven}
\end{figure*}

\subsection{Analysis of Disease-gene associations}
The identified 141 genes involved in four significant pathways (cellular process, signalling transduction, viral and neurodegenerative) are further screened by looking into their association with \covid related disease. We particularly focused on three highly influential diseases during \covid, namely cardiovascular, respiratory tract~\cite{wu2020new,clerkin2020covid,konturek2020covid,agapito2018dietos} and immune system disease~\cite{melenotte2020immune,chowdhury2020immune}. To obtain disease-gene association, we used DisGeNET database~\cite{pinero2020disgenet} and selected $CURETED$ source only. We found a total of 64 proteins (out of 141) playing roles in various diseases such as \emph{Asthma, Pneumonitis, Pneumonia, Influenza, Lung diseases, Cardiomyopathies, Coronary, Arteriosclerosis, Coronary Artery Disease, Heart failure, HIV Infections} etc.(\emph{Supplementary-C}). We compared proteins involved in all three disease categories and individual disease in each category (Figure~\ref{fig:ven}). A total of 119, 37, and 48 unique diseases, and 44, 17, and 24 distinct proteins are associated with the Cardiovascular, Respiratory, and Immune system disease category, respectively. Interestingly, we found a few proteins that are associated with all three disease categories ( AREG, CAV1, IFIH1, PARP1, PLAU, TGFB1, ATM, B2M, DDX58, ENO1, HSPA5, PRKDC, STAT6, TGFBR1, TGFBR2). The top few proteins with ten or more disease associations are  PLAU(59), TGFB1(29), CAV1(17), PARP1(17), TGFBR2(13), ATP2A2(11), AREG(10), FASN(10), IFIH1(10), ITGB1(10). The list of all 64 proteins and their associated quantitative parameters (degree, disease count (out of 204), disease type count (out of 3), and pathway count(out of 31) are presented in Table~\ref{tab:all}.

\subsection{Viral proteins targeting key proteins}
We then looked into source network (Figure~\ref{fig:data}) to identify the viral proteins that are targeting our selected 65 disease associated proteins. We found 25 \sars proteins that  are interacting with 65 proteins. Among 25 \sars viral proteins,  eight are accessory proteins (Orf3a, Orf7b, Orf6, Orf7a, Orf7b, Orf8, Orf9b, Orf10), four structural proteins (E,M,N,S) and thirteen non-structural poly-proteins (nsp1, nsp10, nsp12, nsp13, nsp14, nsp2, nsp3, nsp4, nsp5, nsp6, nsp7, nsp8, nsp9). It is observed that several host proteins are interacting with single viral protein. Very few host proteins are interacting with more than one viral proteins. The viral protein Orf7b exhibits the maximum number of target host proteins followed by Orf3a and M protein. Further, five host proteins are found to be common both in Orf3a and Orf7b. 

We look further for any other viruses that are targeting our 65 host proteins. We mine VirusMINT~\cite{chatr2009virusmint}, a virus-host association database, to find the other related viral diseases. We found that the majority of the highlighted host proteins are also targeted by  \emph{Hepatitis C virus genotype 1b, Poliovirus Type 1, Human herpesvirus 1, Human papillomavirus type 16 \& 31, Simian virus 40, Sendai virus,  Human adenovirus 5 \& 12, Epstein-Barr virus, Human SARS coronavirus
Bovine papillomavirus type 1,} and \emph{Epstein-Barr virus} (Table~\ref{tab:all}). These proteins might be highly essential and need to put uttermost importance on developing host-directed antiviral therapies for~\covid.

\begin{figure*}[!ht]
\centering
\includegraphics[scale=.99]{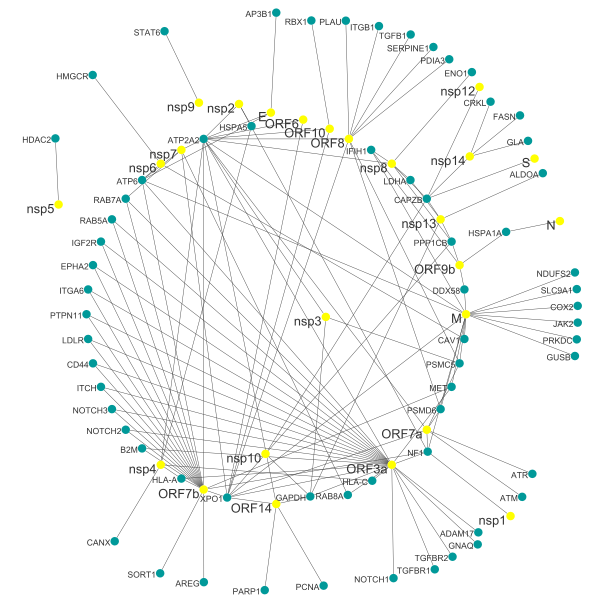}\label{fig:hpvp}
\caption{The interaction network represents the most influential host protein and viral protein. The network is consisting of sixty-four human proteins interacting with twenty-five \sars viral proteins. The yellow colour represents the viral protein in the network, whereas the green one represents the host protein.}
\label{fig:mut-dist-per-variant}
\end{figure*}

\begin{table*}[!ht]
\scriptsize 
\centering
\caption{The table presents sixty-four genes/proteins. Each protein is represented with a degree in PPI (Figure~\ref{fig:ppi}), disease count (out of 204), disease type count (out of 3), and pathway count (out of 31). Some of the proteins have other known virus target are also reported.} \label{tab:all}
\begin{tabular}{lllp{2.8cm}p{1cm}p{6.5cm}} \hline 
\textbf{Gene}     & \textbf{Degree} & \textbf{\#Pathway} & \textbf{Disease category}   & \textbf{\#Disease count} & \textbf{Known target virus}    \\ \hline 
ADAM17   & 10     & 2        & Immune                                & 1              &   \\
ALDOA    & 26     & 1        & Cardiovascular                        & 1              & \\
AP3B1    & 17     & 1        & Respiratory                           & 2              & \\
AREG     & 20     & 2        & Respiratory, Cardiovascular,   Immune & 10             &  \\
ATM      & 29     & 6        & Cardiovascular, Immune                & 8              &\\
ATP2A2   & 10     & 1        & Cardiovascular                        & 11             &\\
ATP6     & 19     & 5        & Cardiovascular                        & 1              & Human SARS coronavirus,  Bovine   papillomavirus type 1,  Human   papillomavirus type 16  \\
ATR      & 17     & 5        & Respiratory                           & 2              & Human adenovirus 5   \\
B2M      & 25     & 4        & Cardiovascular, Immune                & 4              & Hepatitis C virus genotype 1b (isolate Con1)  \\
CANX     & 31     & 2        & Cardiovascular                        & 1              &  \\
CAPZB    & 34     & 1        & Cardiovascular                        & 1              &   \\
CAV1     & 22     & 2        & Respiratory, Cardiovascular,   Immune & 17             & Poliovirus type 1 (strain Sabin)   \\
CD44     & 19     & 1        & Immune                                & 1              &  \\
COX2     & 9      & 3        & Cardiovascular                        & 1              &  \\
CRKL     & 11     & 5        & Cardiovascular                        & 4              &  \\
DDX58    & 10     & 6        & Respiratory, Cardiovascular           & 2              & \\
ENO1     & 13     & 1        & Cardiovascular, Immune                & 2              &  \\
EPHA2    & 9      & 2        & Cardiovascular                        & 5              & \\
FASN     & 9      & 1        & Cardiovascular                        & 10             &  \\
GAPDH    & 23     & 2        & Cardiovascular                        & 1              & Hepatitis C virus genotype 1b (isolate Con1),  Epstein-Barr virus (strain GD1)  \\
GLA      & 16     & 1        & Cardiovascular                        & 5              &  \\
GNAQ     & 14     & 5        & Cardiovascular                        & 1              & \\
GUSB     & 16     & 1        & Immune                                & 1              &\\
HDAC2    & 12     & 5        & Respiratory                           & 2              & Human herpesvirus 1 (strain 17),    Human papillomavirus type 16,    Human papillomavirus type 31 \\
HLA-A    & 15     & 8        & Immune                                & 5              & Epstein-Barr virus (strain GD1),    Human papillomavirus type 16                                  \\
HLA-C    & 14     & 8        & Immune                                & 7              &                                                                                                   \\
HMGCR    & 11     & 1        & Immune                                & 4              &                                                                                                   \\
HSPA1A   & 30     & 5        & Cardiovascular                        & 1              & Epstein-Barr virus (strain GD1)                                                                   \\
HSPA5    & 46     & 1        & Respiratory, Cardiovascular           & 3              & Epstein-Barr virus (strain GD1)                                                                   \\
IFIH1    & 10     & 3        & Respiratory, Cardiovascular,   Immune & 10             & Sendai virus (strain Fushimi)                                                                     \\
IGF2R    & 31     & 2        & Respiratory                           & 1              &                                                                                                   \\
ITCH     & 46     & 1        & Immune                                & 1              & Epstein-Barr virus (strain B95-8)                                                                 \\
ITGA6    & 10     & 3        & Immune                                & 1              &                                                                                                   \\
ITGB1    & 29     & 5        & Cardiovascular                        & 10             & Hepatitis C virus genotype 1b   (isolate Con1)                                                    \\
JAK2     & 22     & 2        & Cardiovascular                        & 6              &                                                                                                   \\
LDHA     & 14     & 1        & Cardiovascular                        & 4              &                                                                                                   \\
LDLR     & 20     & 2        & Cardiovascular                        & 4              &                                                                                                   \\
MET      & 18     & 4        & Respiratory                           & 1              &                                                                                                   \\
NDUFS2   & 20     & 3        & Cardiovascular                        & 3              &                                                                                                   \\
NF1      & 16     & 2        & Cardiovascular                        & 3              &                                                                                                   \\
NOTCH1   & 23     & 3        & Cardiovascular                        & 4              & Hepatitis C virus genotype 1b   (isolate Con1)                                                    \\
NOTCH2   & 11     & 2        & Cardiovascular                        & 1              &                                                                                                   \\
NOTCH3   & 15     & 3        & Cardiovascular                        & 2              &                                                                                                   \\
PARP1    & 14     & 1        & Respiratory, Cardiovascular,   Immune & 17             & Human herpesvirus 1 (strain 17)                                                                   \\
PCNA     & 25     & 3        & Immune                                & 1              & Human herpesvirus 1 (strain 17)                                                                   \\
PDIA3    & 14     & 3        & Cardiovascular                        & 1              &                                                                                                   \\
PLAU     & 24     & 1        & Respiratory, Cardiovascular,   Immune & 59             &                                                                                                   \\
PPP1CB   & 14     & 4        & Cardiovascular                        & 2              &                                                                                                   \\
PRKDC    & 15     & 1        & Respiratory, Immune                   & 3              & Human herpesvirus 1 (strain 17)                                                                   \\
PSMC5    & 34     & 1        & Immune                                & 7              & Human adenovirus 5, Human   adenovirus 12,  Simian virus 40                                       \\
PSMD6    & 24     & 1        & Immune                                & 2              &                                                                                                   \\
PTPN11   & 22     & 1        & Cardiovascular                        & 6              &                                                                                                   \\
RAB5A    & 40     & 3        & Cardiovascular                        & 1              &                                                                                                   \\
RAB7A    & 41     & 2        & Cardiovascular                        & 1              &                                                                                                   \\
RAB8A    & 40     & 3        & Immune                                & 1              &                                                                                                   \\
RBX1     & 58     & 4        & Immune                                & 5              &                                                                                                   \\
SERPINE1 & 16     & 4        & Cardiovascular                        & 8              &                                                                                                   \\
SLC9A1   & 12     & 2        & Cardiovascular                        & 7              &                                                                                                   \\
SORT1    & 14     & 1        & Cardiovascular                        & 3              &                                                                                                   \\
STAT6    & 11     & 1        & Respiratory, Immune                   & 4              &                                                                                                   \\
TGFB1    & 29     & 7        & Respiratory, Cardiovascular,   Immune & 29             & Hepatitis C virus genotype 1b   (isolate Con1)                                                    \\
TGFBR1   & 17     & 9        & Respiratory, Cardiovascular           & 6              &                                                                                                   \\
TGFBR2   & 16     & 8        & Respiratory, Cardiovascular           & 13             &                                                                                                   \\
XPO1     & 25     & 2        & Cardiovascular                        & 2              &   \\   \hline    
\end{tabular}
\end{table*}

\section{Conclusion}

In this study, we have analysed  human host protein-protein interaction network during the \sars infection. We identified a set of proteins, including RBX1, HSPA5, ITCH, RAB7A, RAB5A, RAB8A, PSMC5, CAPZB, CANX, IGF2R, HSPA1A, which might influence the whole PPI network. These proteins were enriched for the following processes: cellular process, signalling, and neurodegenerative disease pathways as these pathways are known to be highly infectious for disease pathogenesis during \covid. Finally, we have found 64 potential/key \sars interacting human host proteins connected with respiratory, cardiovascular, and immune system disease. Many of them are known to target different other viruses and may be highly important for therapeutics treatment of \covid viral disease. We strongly believe that the highlighted key proteins are an extremely promising target, which might play a crucial role during \covid disease progression.

\bibliographystyle{elsarticle-num-names}
\bibliography{sample.bib}

\end{document}